# Comparative analysis of full-field OCT and optical transmission tomography

SAMER ALHADDAD,[1] OLIVIER THOUVENIN,[1] MARTINE BOCCARA,[2] CLAUDE BOCCARA,[1] AND VIACHESLAV MAZLIN[1,*]

[1]*Institut Langevin, ESPCI Paris, PSL University, CNRS, 1 rue Jussieu, 75005 Paris, France*
[2]*Institut de Systématique, Evolution, Biodiversité (ISYEB), Muséum National d'Histoire Naturelle, Sorbonne Université, EPHE, UA, CNRS ; CP 50, 57 rue Cuvier, 75005 Paris, France*
*\*mazlin.slava@gmail.com*

**Abstract:** This work compares two tomographic imaging technologies, time-domain full-field optical coherence tomography (FFOCT) working in reflection and optical transmission tomography (OTT), using a new optical setup that combines both. We show that, due to forward-scattering properties, the axial sectioning and contrast in OTT can be optimized by tuning illumination. The influence of sample scattering and thickness are discussed. We illustrate the comparison of the two methods in static (morphology) and dynamic (metabolic contrast) regimes using cell cultures, tissues and entire organisms emphasizing the advantages of both approaches.

## 1. Introduction

Transmission bright field imaging is indispensable for observing morphology of biological samples. At the origin of image formation is the phenomenon of light interference between the transmitted and scattered waves [1,2]. The classical work of Zernike [3] demonstrated the ability to control image contrast by varying the relative phase between the waves. Further advancements allowed to perform quantitative phase measurements [4]. These techniques excel at imaging semitransparent biological samples. However, analyzing thick and scattering samples is challenging since the in-focus structures of interest can be hindered by the light coming from out-of-focus sample layers. This can decrease the image quality and result in a loss of information, making it difficult to analyze the sample accurately.

Another class of approaches that are commonly used for imaging thick samples are based on interferometry of backscattered light. Optical coherence tomography (OCT) is one such technique that relies on broadband illumination spectrum, an interferometer, and a spectrometer to reconstruct tomographic views of the sample without out-of-focus light [5,6]. A specific subtype of OCT, known as time-domain full-field OCT (FFOCT) takes advantage of 2D cameras to obtain en face images [7,8]. The tomographic images are then reconstructed from a sequence of camera frames, each containing a distinct light interference pattern or optical phase. FFOCT has been applied to various imaging problems, such as cancer biopsy analysis [9], visualization of organoid development [10], fingerprint identification [11], and ophthalmology [12–15].

Recently, we introduced a transmission-based optical sectioning approach, called optical transmission tomography (OTT) [16]. Similar to FFOCT, OTT reconstructs tomographic images from a sequence of phase-shifted camera frames. Compared to OCT methods, OTT has its axial resolution, determined by the numerical aperture. In addition to the static tomographic views, the above phase-based methods can produce color-coded images of intrinsic metabolic cell dynamics.

The aim of this paper is to provide a comparative analysis of FFOCT and OTT and to illustrate the differences between these two approaches using a few model cases. To achieve this goal, we combined FFOCT and OTT in a single microscope to image the same field of view with both modalities. We first briefly describe the principles of both techniques and then demonstrate that, due to forward scattering properties, axial resolution in OTT depends on the filling of the objective pupil aperture and is object dependent. Taking advantage of this effect, we propose adding a diffusive layer between the illumination and the object in OTT. We discuss a newly discovered compromise between sectioning and contrast that is specific to OTT, as well as the influence of thickness of samples observable with this modality. Finally, we use FFOCT-OTT microscope to compare tomographic images of both modalities acquired in transparent zebrafish embryos and different murine organs, showing that the accessible contrast can be significantly different.

## 2. Methods

### 2.1 Combined FFOCT-OTT system

FFOCT is composed of a Linnik interferometer with identical microscope objectives (MO) (water-immersion 40X Nikon CFI APO, 0.80 numerical aperture (NA), 3.5 mm working distance (WD)) in the sample and reference interferometric arms. Illumination is provided by an incoherent 660 nm light-emitting diode (LED) (M660L3, Thorlabs, USA) to avoid light speckle and cross-talk artifacts. The LED has low temporal coherence with a spectral bandwidth of 25 nm, which determines axial resolution, similar to conventional OCT. The reference arm is equipped with a mirror (silicon wafer, 40% reflectivity at 660 nm) mounted on a piezoelectric motorized stage (STr-25/150/6, Piezomechanik GmbH, Germany). The piezo stage modulates the mirror position in time, controlling the optical phase delay (0 or $\pi$) between the sample and reference light beams. The camera (Q-2A750-CXP, Adimec, Netherlands) sequentially captures two phase-shifted frames. Subtracting one frame from another rejects the non-interfering light from the rest of the sample and produces an en face tomographic view of the slice of interest.

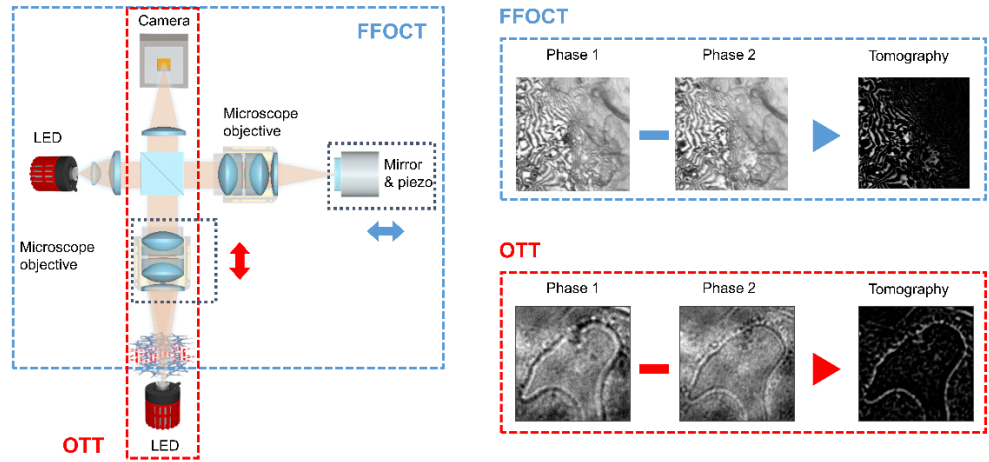

Fig. 1. Combined FFOCT-OTT setup and principles of tomographic reconstruction.

OTT is integrated into FFOCT through an additional LED (M660L3, Thorlabs, USA) located underneath the sample (Fig. 1). OTT relies on a common-path transmission interferometer geometry, where light scattered by the sample interferes with light transmitted through the sample. Although the two beams propagate along the same optical path, they differ in divergences, with the scattered wave filling a larger portion of the objective's numerical aperture. One can control the phase delay between the two beams using the Gouy effect - a $\pi$ optical phase shift that a light wave exhibits when passing through the optical focus. The Gouy phase shift appear as a fundamental property of the focused Gaussian beam model [17]. By axially moving the objective lens, we change the optical phase of the light beam scattered from the in-focus slice of the sample. Similar to FFOCT, sequentially acquired phase-shifted frames can be subtracted from one another to reject the out-of-focus light that is not affected by the Gouy phase shift and reconstruct the en face tomographic view.

FFOCT and OTT acquisitions are not simultaneous; when one LED is working, the other is disabled. A comparison of FFOCT and OTT is summarized in Table 1.

Table 1. Comparative table of FFOCT and OTT

|  | FFOCT | OTT |
|---|---|---|
| Illumination wavelength | 660 ±12nm | 660±12nm |
| Field of view | 260μm × 260μm | 260μm × 260μm |
| Lateral resolution | $\lambda/(2 \cdot NA)$ (0.5μm) | $\lambda/(NA_c + NA_i)$ (0.5μm – 0.8μm) |
| Axial Resolution / Depth of field | $1.4 \cdot \lambda^2/(\pi \cdot n \cdot \Delta\lambda)$ (6μm) | $2 \cdot \lambda \cdot (n^2 - NA^2)/(\pi \cdot n \cdot NA)$ (0.5μm - 50μm depending on the illumination and sample) |
| Tomographic reconstruction | phase shifting by piezo modulation of mirror in reference arm | phase shifting by piezo modulation of microscope objective in both arms (Gouy effect) |
| Contrast | back-scattered | forward-scattered |

*$\lambda$ is the wavelength of light, $NA_c$ is numerical aperture of microscope objective, $NA_i$ is numerical aperture of illumination, n is the refractive index of the sample

## 2.2 Biological samples

Zebrafish larvae are a reliable model organism due to their transparency, small size, fast development, genetic tractability and similarity to human biology. Experiments on zebrafish larvae were performed at 5 days post fertilization (dpf) following procedures approved by the Institutional Ethics Committee Darwin in the Institut du Cerveau (ICM). Zebrafish larvae grew in an incubator at 28.5°C until 5 dpf. Shortly before the experiment, the larvae were laterally mounted in 1.5% low melting point agarose and paralyzed by injecting 0.5 nl of 0.5 mM α-bungarotoxin in the ventral axial musculature. The experiments were performed at room temperature in a physiological solution. Murine organs were obtained from the partner research institution Institut de la Vision (Paris) as recuperated waste tissue from the unrelated experiment. Imaging was performed within few hours after dissection.

## 3. Experimental results

### 3.1 Axial resolution in FFOCT and OTT

The unique feature of OCT techniques in decoupling of axial resolution from the NA of the imaging optics is not applicable to OTT. However, the axial resolution in OTT can be controlled not only via NA of detection optics but also via illumination NA. To investigate this effect, we performed two tests: first, we closed an aperture directly above the LED to produce an effective light source of small size and reduce the NA of illumination; second, we covered the LED with a scattering medium (16 parafilm layers about 2 mm in total thickness), which extended the light source and increased the illumination NA. In both tests the LED was located 1 cm from the sample imaging plane, and the 5 dpf zebrafish tail was used as a sample. The direct transmission images from these configurations are compared in Fig. 2 and Visualization 1. As expected, the case of illumination through a scattering medium showed a significant improvement in axial resolution and clearly distinguishable layers. In comparison, the conventional illumination the light source is several cm away from the objective, which limits the filling NA and axial sectioning. In the opposite case of aperture-limited illumination, sample layers remain sharp across the large axial focus beyond 50 μm.

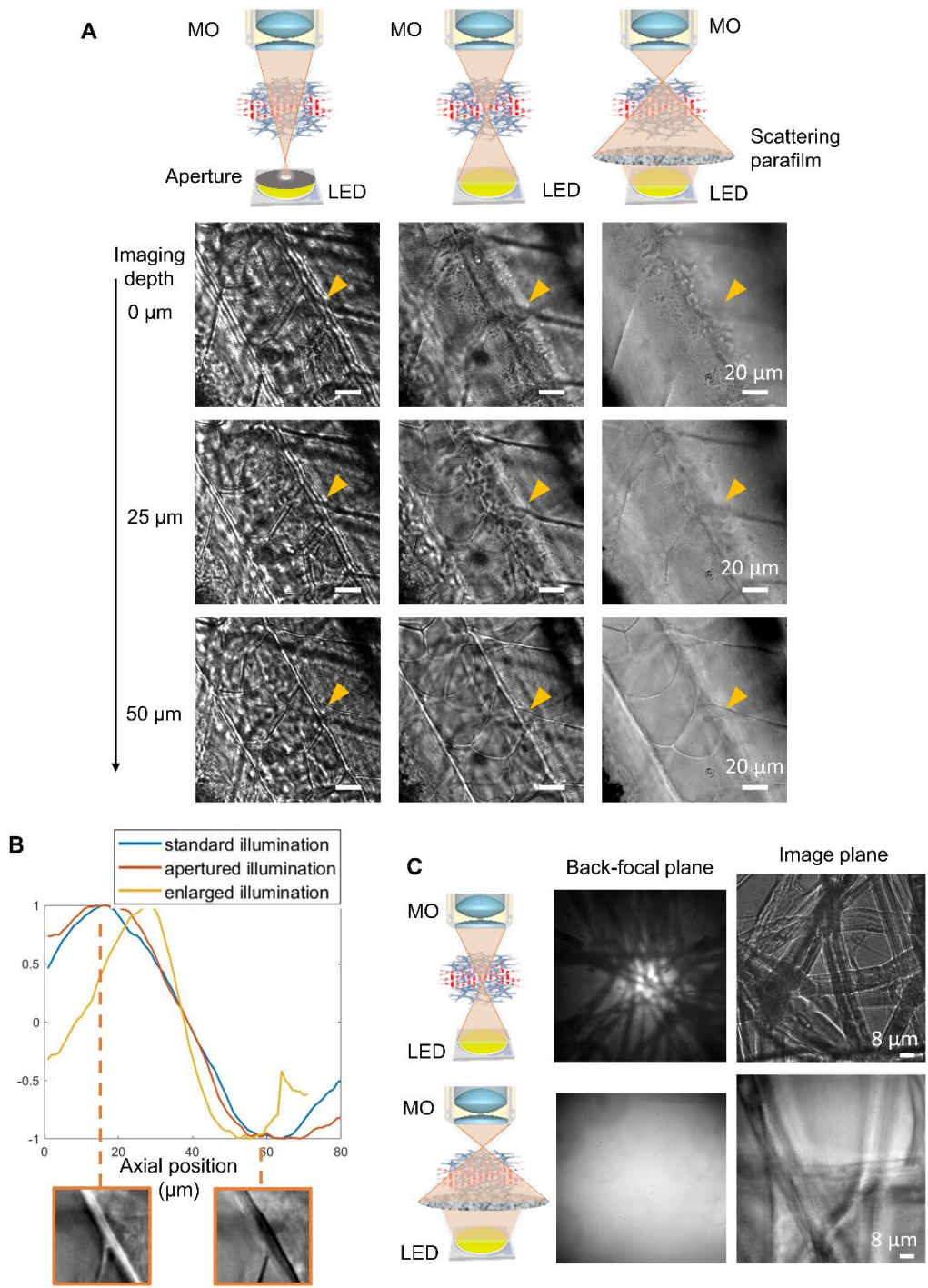

Fig. 2. Three configurations of illumination in OTT (small, standard and extended source) and their effects on axial sectioning and contrast (A, B). Through depth sequences are available in Visualization 1. Graphs in B are calculated on the selected structure shown with yellow arrows in A. Extended source fills larger portion of the detection NA producing better optical sectioning. Increase in resolution is accompanied by the decrease in interference contrast, as transmitted and scattered waves become co-aligned and exhibit the same phase (C). In order to fill the detection NA the scattering parafilm was placed directly beneath the sample.

Interestingly, we observed that the improvement in axial resolution in OTT comes at the cost of a decrease in contrast. To explain this inverse relationship, we used a separate OTT system equipped with a microscope objective (PA100X-INF-IRIS, Amscope) that had an accessible back-focal plane, and we imaged fibers of the optical cleaning tissue. Our findings showed that the interference contrast in the image plane completely vanishes when the illumination NA fills the detection NA (as illustrated in Fig. 2C). This case corresponds to transmitted and scattered beams having similar angular divergences and similar phases, as a result no interference phenomena can be observed. In the intermediate cases, the more collimated is the illumination beam, the larger is the angular separation between the transmitted and scattered beams, leading to a larger interference contrast in the image.

It is important to note similarities between imaging with the scattering layer and imaging thick samples. Indeed, the sample thickness below the imagine plane acts as a scattering parafilm between the light source and the sample. The thicker is the sample the broader is the secondary illumination of the imagine plane and the more filled the numerical aperture leading to reduction of interferometric contrast. The best resolution-contrast tradeoff depends on the sample thickness and degree of scattering.

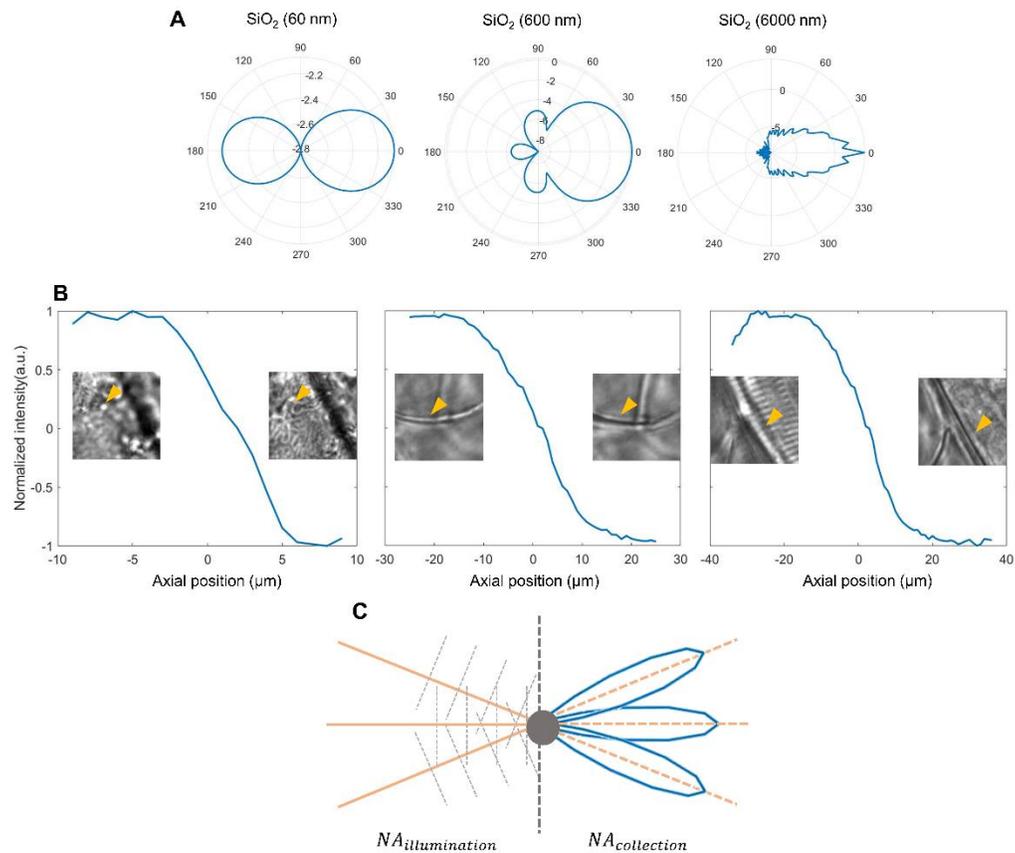

Fig. 3. Effects of Mie scattering in OTT. A, Mie scattering intensity from $SiO_2$ particles (n = 1.45) of different sizes in water (n = 1.33) presented in logarithmic scale. B, Gouy phase profiles for scatterers of different sizes in zebrafish. Embedded OTT images are 25 μm × 25 μm. C, Collection NA is affected by both illumination NA and angular divergence of scattered light from the sample structures. Blue curves correspond to 1 μm $SiO_2$ particles.

Finally, we would like to illustrate that the axial resolution in OTT is dependent on the nature of the scattering objects (Fig. 3). If the scatterer is small compared to the wavelength

(diameter <1/10λ) light exhibits isotropic radiation (Rayleigh regime). However, if the scatterer is larger with size close to the wavelength, the light will be mostly forward scattered, in the same direction as the incident light (Mie scattering regime). For instance, for $SiO_2$ particles of 60, 600 and 6000 nm in diameter illuminated by a 660 nm plane wave the average cosine of this angle is given by the anisotropic scattering parameter g=0.025, 0.87 and 0.95 respectively (Fig. 3A) [18]. Similar principle is seen in zebrafish sample (Fig. 3B). Biological structures are predominantly forward scatterers [19], nevertheless the degree of angular divergence can differ from structure to structure. Small structures exhibit fast Gouy phase shift across the optical focus (~ 10 µm), as the divergently scattered light wave fills the detection NA. On the other hand, larger structures scatter light in the forward direction, underfill the NA and are effectively seen with larger depth of field (~ 40 µm).

In reality the detection NA depends on both the illumination NA and on the angular distribution of the scattering from the objects (Fig. 3C).

### 3.2 Tomographic image comparison

Below we provide comparison of tomographic FFOCT and OTT images obtained by subtraction of two π-phase-shifted camera frames. During OTT acquisition the scattering parafilm was placed between the LED and the sample to improve axial resolution. Fig. 4 and Fig. 5 present results from zebrafish larvae tail and mouse organs (brain, peritoneum, heart), respectively.

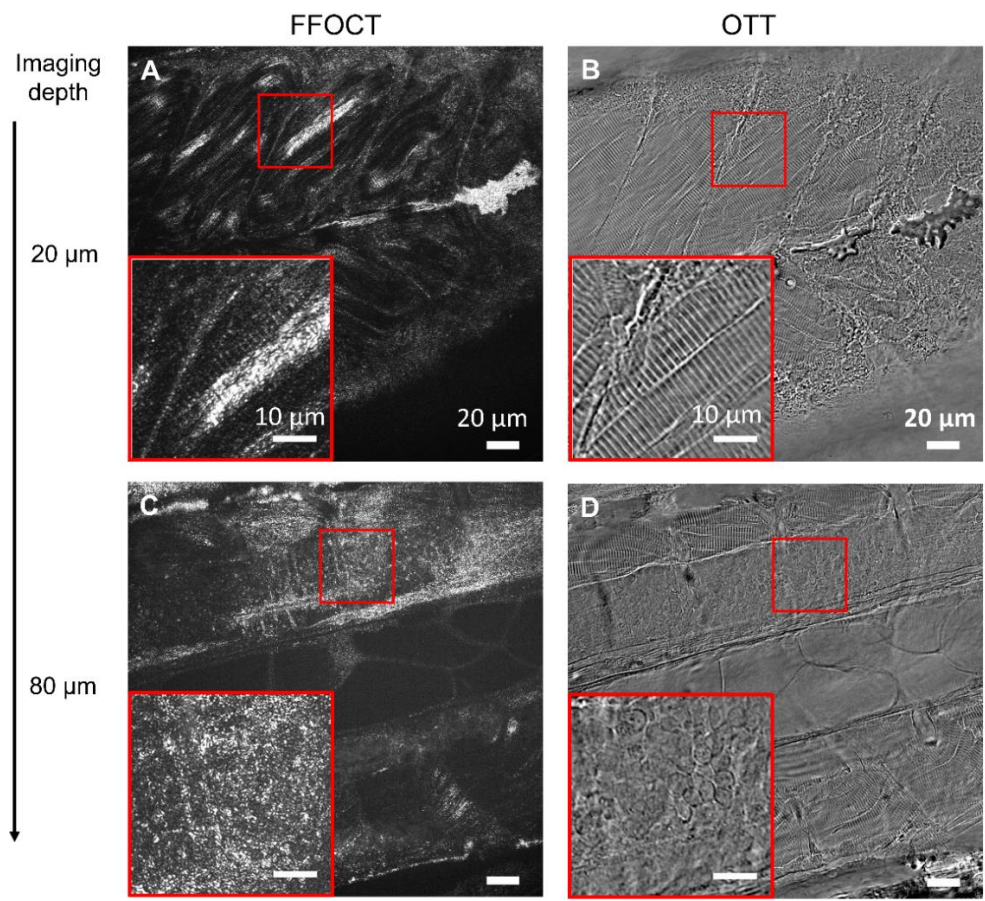

Fig. 4. Tomographic FFOCT (40 averages) and OTT (no averaging) images of zebrafish larvae tail. FFOCT has finer optical sectioning, however view on small cellular mosaics can be obstructed by the residual interference fringe artifacts.

FFOCT demonstrates higher optical sectioning and can undoubtedly provide superior volumetric images than OTT. On another hand, in single en face views OTT presents a number of advantages over FFOCT. First, FFOCT required averaging of 40 tomographic images to improve signal-to-noise ratio to the level of a single OTT image. This is due to both the smaller scattering volume of inside the fine section in FFOCT and higher forward scattering in transmission-based OTT. Secondly, the fine resolution of FFOCT can occasionally make the image analysis difficult because only parts of the sample structures are visible. This is illustrated in Figs. 4(A,B), where lower axial resolution of OTT allows it to resolve the full length of somites contrary to FFOCT. Lastly, FFOCT is frequently affected by the residual interference fringe artifacts that may prohibit resolving fine cellular mosaics at the resolution limit (Fig. 4(C,D)). In general, FFOCT and OTT were able to successfully resolve different zebrafish structures including notochord, blood vessels, skin layer, cells from spinal cord and some of their axons.

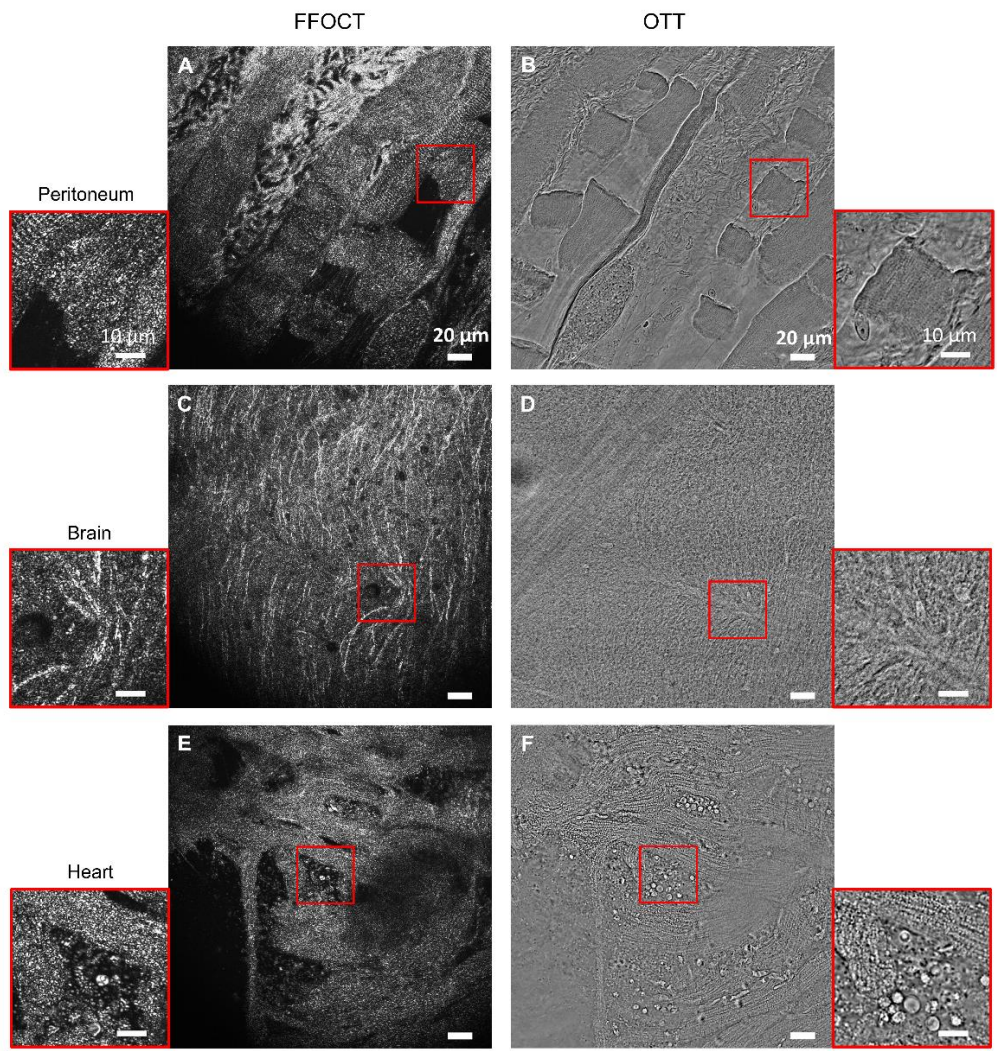

Fig. 5. Tomographic FFOCT (40 averages) and OTT (no averaging) images of mouse organs. All samples were imaged at 30 µm depth. Through depth sequences for brain and heart are available as Visualizations 2 and 3.

Mouse structures can be more complex to image than that of transparent zebrafish embryo because of the scattering properties and thickness of the tissues. For these measurements we cut a thin layer of the sample of interest and we flatten the sample by squeezing it between the lower coverslip and a thin optically transparent pellicle membrane. Both FFOCT and OTT resolved fine structures across large depths (see Visualizations 2 and 3). Particularly, in the brain the axons of cortical neurons were highlighted in FFOCT, while the neuron cell bodies appeared as dark disks as previously reported [20,21]. In contrast, the axons in OTT lacked contrast, but other cell structures were readily detectable.

*3.3 Dynamic image comparison*

FFOCT and OTT can extract two types of physiological dynamics: blood flow [15,22] and metabolic cell activity [10,23].

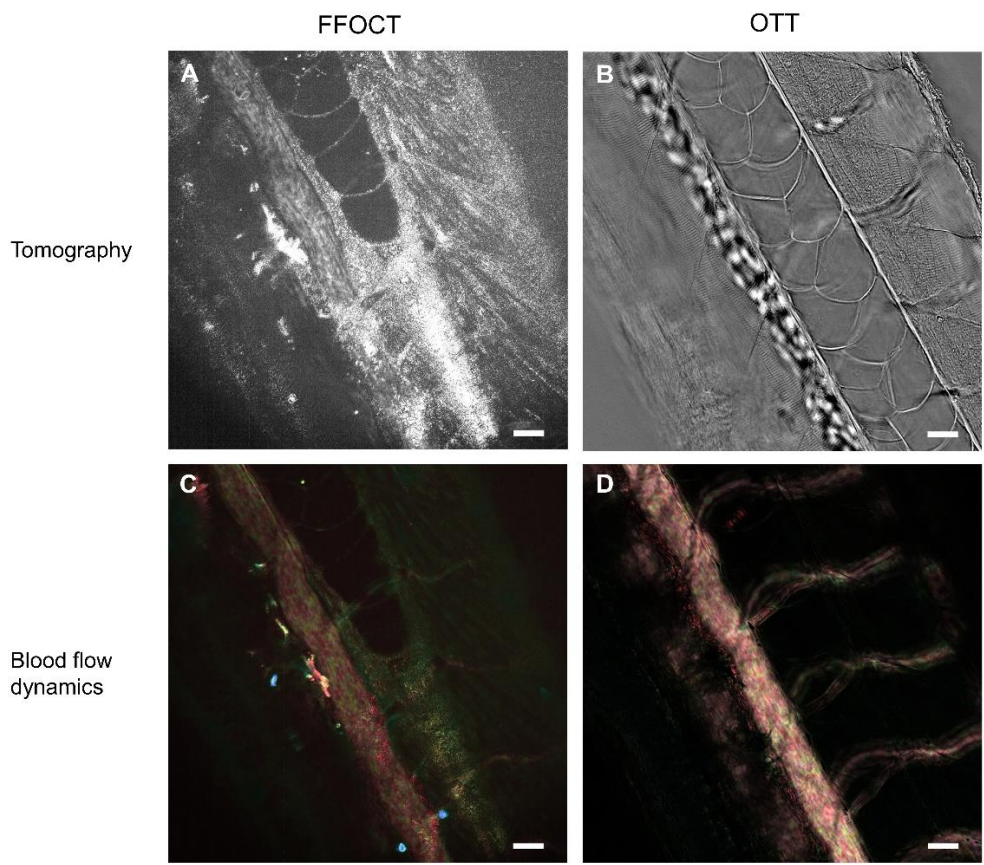

Fig. 6. Blood flow dynamics in FFOCT and OTT in zebrafish embryo. Scale bars are 20 μm.

As the location of erythrocytes changes from one camera frame to another, tomographic images will show them even, if they propagate above or below the optical section [24]. OTT underfills the collection NA, therefore it shows the entire blood flow pathway (Fig. 6). FFOCT on the contrary uses the entire NA, thus visualizes fewer vessels within the shallower depth of field.

In the second experiment (Fig. 7), we looked at metabolic cell dynamics of mouse liver in FFOCT and OTT. Dynamic experiments were performed with two separate FFOCT and OTT setups. Color HSV dynamic images were reconstructed from 256 sequential camera frames with the Hue, Saturation, Value channels encoding signal's central fluctuation frequency, frequency range and fluctuation amplitude, respectively [16]. Direct camera frame in FFOCT captures little signals in reflectance, while transmission-based OTT shows many large particles with Gouy phase interference contrast. These particles are likely glycogen granules used to store fat in hepatocytes. Tomographic images are rather homogeneous in FFOCT with visible nuclei, while in OTT tomography shows the granules. In dynamic images, FFOCT shows a rather homogeneous signal in cell cytoplasm and faster signal in nuclei, the same signal being visible in dynamic OTT for cells further from the sample surface. Further studies with identical field of views in FFOCT and OTT would be valuable for future quantitative comparison.

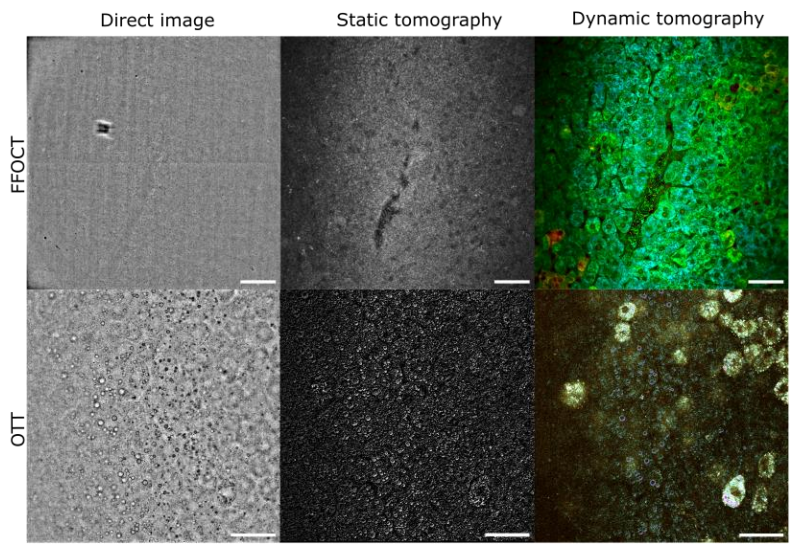

Fig. 7. Comparison of metabolic cell dynamics in mouse liver. Scale bars are 50 µm.

### 3.4 FFOCT and OTT in cell cultures

Reflection-based FFOCT is known to produce sharp interference fringe artifacts from the glass slide underneath the cell cultures. OTT, on another hand, produces clean images thanks to its transmission geometry (Fig. 8).

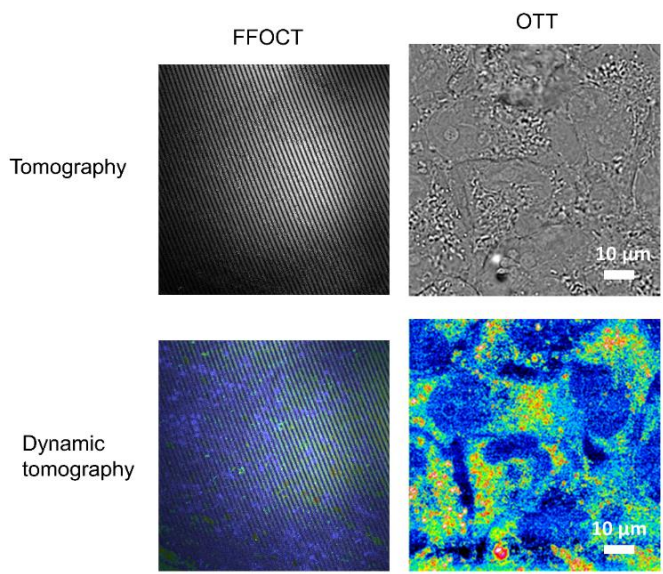

Fig. 8. OTT is immune to interference fringe artifacts that prohibit FFOCT from imaging cell cultures on a glass slide. The example images are taken with COS cancer cells.

## 4. Discussion

We conducted a comparison study of two optical tomographic techniques, namely FFOCT and the more recent OTT using the combined experimental set-up. FFOCT provides exceptional axial resolution determined by coherent gating with a light source. In OTT the optical sectioning is fixed with the numerical aperture. However, we can tune it by changing the numerical aperture of illumination. That can be done with the aperture and scattering medium without changing the light source. Unfortunately, increase in resolution is accompanied by the decrease in contrast, due to filling of the objective's pupil and minimization of phase difference between the interfering transmitted and scattered light waves. The same effect is seen when imaging thick scattering tissues as the light scattered from them underneath the image plane is acting as an extended secondary source that fills the objective's pupil. Therefore, high optical sectioning of order of depth-of-field can be practically achieved in OTT only in transparent or thin samples. Other factors that reduce sectioning include lower quality of the axial PSF due to additional wings from the Gouy phase curve [16] and dependence of axial PSF on the scatterer size (Mie scattering).

While axial sectioning advantage of FFOCT would be obvious in volumetric images, the single en face views of OTT are frequently easier to interpret. For example, each high-resolution slice of FFOCT shows only part of the structure of interest, while OTT shows the full structure. Additionally, OTT is immune to residual interference artifacts of FFOCT that prohibit imaging of fine cellular mosaic at the resolution limit and also make imaging of cell cultures on a glass slide impossible.

Depending on the physical profile of the structure its contrast can be better in transmission-based OTT or reflection-based FFOCT. For example, thin axons of cortical neurons are clearly seen in FFOCT while round cellular bodies are better resolved in OTT.

In terms of signal FFOCT needs to average about 40 frames to match with OTT. That is due to: 1) the double optical arm geometry of FFOCT with intense non-interfering background light, which reduces the useful full well capacity of the camera, 2) smaller number of scatterers in the finer sectioning volume, 3) predominantly forward scattering of biological tissues. Therefore, OTT images can be acquired faster with shorter exposure time. That being said, in thick samples this rule is reversed and OTT requires long exposure times over 100 ms to saturate the camera.

Both FFOCT and OTT can follow the blood flow and metabolic cell dynamics.

Common-path optical design of OTT is particular advantageous in terms of robustness, cost and reduced sensitivity to external vibration.

**Funding.** Agence Nationale de la Recherche (project 'CERES', project 'DEEPINCELL', ANR-10-IDEX-0001-02 PSL).

**Acknowledgements.** We warmly thank Martin Carbó-Tano for preparation of zebrafish larvae and Marie Darche for providing the murine organs.

**Disclosures.** SA: none, OT: none, MB: none, CB: none, VM: none.

**Data availability.** Full resolution images generated in this paper can be obtained from the authors upon request.